\documentclass[12pt]{article}
\usepackage[dvips]{graphicx}
\usepackage{amsfonts,amssymb,amsthm,amstext,amscd}

\newcommand{\beqa}{\begin{eqnarray}}
\newcommand{\eeqa}{\end{eqnarray}}

\def\rhne{r_0}
\def\rhor{r_h}
\def\epshne{\epsilon_h}
\def\Phihne{\Phi_h}

\def\eqref#1{(\ref{#1})}
\def\bardens{n_b}
\def\quarkdens{n_q}
\def\d{{\rm d}}

\newcommand{\be}{\begin{equation}}
\newcommand{\ee}{\end{equation}}
\newcommand{\beq}{\begin{equation}}
\newcommand{\eeq}{\end{equation}}
\newcommand{\bea}{\begin{eqnarray}}
\newcommand{\eea}{\end{eqnarray}}
\newcommand{\nn}{\nonumber}

\newcommand{\bear}{\begin{eqnarray}}
\newcommand{\eear}{\end{eqnarray}}
\begin{document}
\baselineskip=15.5pt
\pagestyle{plain}
\setcounter{page}{1}
%--------+---------+---------+---------+---------+---------+---------+
%Body

\def\r{\rho}
\def\CC{{\mathchoice
{\rm C\mkern-8mu\vrule height1.45ex depth-.05ex
width.05em\mkern9mu\kern-.05em}
{\rm C\mkern-8mu\vrule height1.45ex depth-.05ex
width.05em\mkern9mu\kern-.05em}
{\rm C\mkern-8mu\vrule height1ex depth-.07ex
width.035em\mkern9mu\kern-.035em}
{\rm C\mkern-8mu\vrule height.65ex depth-.1ex
width.025em\mkern8mu\kern-.025em}}}

%%%%%%%%%%%%%%%%%%%%%%%%%%%%%%%%%%%%%% Mas and Tarrio definitions %%%%%%%%%%%%%%%%%%%%%%%%%%%%%%%%

\def\cst{c_T}
\def\csx{c_X}
\def\csr{c_R}
\def\css{c_S}
\def\csf{c_F}
\def\csu{c_U}

\def\hxm{{\hat{x}^\mu}}
\def\hxn{{\hat{x}^\nu}}
\def\hr{{\hat{r}}}
\def\ha{{\hat{a}}}
\def\hb{{\hat{b}}}
\def\hc{{\hat{c}}}
\def\hd{{\hat{d}}}
\def\htau{{\hat{\tau}}}
\def\hi{{\hat{x}^i}}
\def\hj{{\hat{j}}}
\def\hK{{\hat{K}}}
\def\hL{{\hat{L}}}
\def\hM{{\hat{M}}}
\def\hN{{\hat{N}}}
\def\d{\partial}
\def\med{\frac{1}{2}}
\def\rstar{r_*}
\def\dent{\tilde \delta}
\def\den{\delta}

% FONTS
\newfont{\namefont}{cmr10}
\newfont{\addfont}{cmti7 scaled 1440}
\newfont{\boldmathfont}{cmbx10}
\newfont{\headfontb}{cmbx10 scaled 1728}
\renewcommand{\theequation}{{\rm\thesection.\arabic{equation}}}
\font\cmss=cmss10 \font\cmsss=cmss10 at 7pt

\par\hfill ITP-UU-10/39
\par\hfill DFTT 27/2010

\begin{center}
{\LARGE{\bf D3-D7 Quark-Gluon Plasmas at Finite Baryon Density}}
\end{center}
\vskip 10pt
\begin{center}
{\large
Francesco Bigazzi ${}^{a}$, Aldo L. Cotrone ${}^{b}$, Javier Mas ${}^{c}$, \\
Daniel Mayerson ${}^{d}$, Javier Tarr\'\i o  ${}^{e}$.}
\end{center}
\vskip 10pt
\begin{center}
\textit{$^a$ Dipartimento di Fisica e Astronomia, Universit\'a di Firenze and INFN Sezione di Firenze; Via G. Sansone 1, I-50019 Sesto Fiorentino (Firenze), Italy.}\\
\textit{$^b$  Dipartimento di Fisica, Universit\'a di Torino and INFN Sezione di Torino; \\Via P. Giuria 1, I-10125 Torino, Italy.}\\
\textit{$^c$ Departamento de  F\'\i sica de Part\'\i culas, Universidade de Santiago de Compostela and Instituto Galego de
F\'\i sica de Altas Enerx\'\i as (IGFAE); E-15782, Santiago de Compostela, Spain.}\\
\textit{$^d$  Institute for theoretical physics, K.U. Leuven;
Celestijnenlaan 200D, B-3001 Leuven,
Belgium.}\\
\textit{$^e$ Institute for Theoretical Physics, Universiteit Utrecht, 3584 CE, Utrecht, The Netherlands.}\\
{\small bigazzi@fi.infn.it, cotrone@to.infn.it, javier.mas@usc.es, drm56@cam.ac.uk, l.j.tarriobarreiro@uu.nl}
\end{center}

\vspace{15pt}

\begin{center}
\textbf{Abstract}
\end{center}
We present the string dual to $SU(N_c)\ {\cal N}=4 $ SYM, coupled to
$N_f$ massless fundamental flavors, at finite temperature and baryon
density. The solution is determined by two dimensionless parameters,
both depending on the 't Hooft coupling $\lambda_h$ at the scale set
by the temperature $T$: $\epshne\sim\lambda_h N_f/N_c$, weighting
the backreaction of the flavor fields and $\tilde\delta\sim
\lambda_h^{-1/2}\bardens/(N_f T^3)$, where $\bardens$ is the baryon
density. For small values of these two parameters the solution is
given analytically up to second order. We study the thermodynamics
of the system in the canonical and grand-canonical ensembles. We
then analyze the energy loss of partons moving through the plasma,
computing the jet quenching parameter and studying its dependence on
the baryon density. Finally, we analyze certain ``optical"
properties of the plasma. The whole setup is generalized to non
abelian strongly coupled plasmas engineered on D3-D7 systems with
D3-branes placed at the tip of a generic singular Calabi-Yau cone.
In all the cases, fundamental matter fields are introduced by means
of homogeneously smeared D7-branes and the flavor symmetry group is
thus a product of abelian factors. \vspace{4pt}{\small \noindent }
\vfill

\newpage
%%%%%%%%%%%%%%%%%%%%%%%%%%%%%%%%%%%%%%%%%%%%%%%%%%%%%%%%%%%%%%%%%%%%%%%%%%%%%%%%%%%%%%%

%%%%%%%%%%%%%%%%%%%%%%%%%%%%%%%%%%%%%%%%%%%%%%%%%%%%%%%%%%%%%%%%%%%%%%%%%%%%%%%%

\section{Introduction and summary}
Heavy ion collision experiments at RHIC and LHC allow us to explore
a relevant corner of the QCD phase diagram (high temperature and
relatively small baryon chemical potential), where the theory is
expected to be deconfined. Both the results collected during the
ten-year run of RHIC \cite{arsene} and the preliminary ones at LHC
\cite{lhc} actually indicate that a quark-gluon ``fireball" is
formed and behaves like a strongly coupled system: a liquid with
very small viscosity over entropy density ratio. Holographic methods
provide interesting tools to analyze these kind of systems. The
simplest and best studied example is the conformal ${\cal N}=4$ SYM
plasma which, unexpectedly, has proven to share some properties with
the QCD one. This fact has stimulated further research works with
the aim of refining this master holographic model, for example by
adding fundamental matter fields. The latter has been performed
mainly in the quenched approximation.

In \cite{D3D7QGP} some of the authors have presented a ten
dimensional black-hole solution dual to the non conformal plasma of
${\cal N}=4$ SYM coupled to $N_f\gg 1$ massless
flavors.\footnote{All the hydrodynamic transport coefficients of the
model were derived in \cite{hydro}.} The latter were introduced by
means of homogeneously smeared D7-branes
\cite{Bigazzi:2005md,cnp,review}, extended along the radial
direction up to the black hole horizon. The smearing reduces the
flavor symmetry group to a product of abelian factors and allows a
simple way to account for the backreaction of the D7-branes and thus
to explore the ``unquenched" regime in the dual field
theory.\footnote{For other holographic studies of thermal unquenched
flavors, see \cite{Casero:2005se}.} The analysis was also
generalized to ${\cal N}=1$ non abelian plasmas engineered on D3-D7
systems with D3-branes placed at the tip of a generic singular
Calabi-Yau cone. In the zero temperature limit, the resulting
backgrounds coincide with those found in
\cite{Benini:2006hh}.\footnote{Other solutions employing the
smearing technique appear in \cite{tutti,benini2008}.}

In the present paper we extend the above construction to include a
finite baryon density (or chemical potential, in the alternative
thermodynamical ensemble) for the flavor fields. Working on this
problem with holographic techniques is especially interesting,
taking also into account that there is no systematic way of dealing
with finite baryon density in strongly coupled QCD (lattice QCD
suffering from the so-called sign problem).

We provide a novel gravity solution, dual to the above class of
flavored plasmas in the planar limit at strong 't Hooft coupling.
While the equations of motion we derive are completely general, the
solution can be given in closed analytic form up to second order in
$\epshne\sim\lambda_h N_f/ N_c$ (where $\lambda_h$ is the 't Hooft
coupling at the temperature $T$ of the plasma) and $\tilde\delta\sim
\lambda_h^{-1/2}\bardens/(N_f T^3)$, where $\bardens$ is the baryon
density.
The gauge theories we focus on become pathological at some UV scale,
developing a Landau pole. This is signaled, for example, by a
running dilaton (accounting for the breaking of conformal invariance
induced by the flavor fields) blowing up at a finite radial value.
Correspondingly, the dual gravity solutions are not reliable close
to that scale. Keeping $\epshne$ small allows both to focus on a
regime where the solutions are reliable and to decouple the IR
physics - which is the regime we focus on - from the pathological UV
behavior.

We also consider the regime of non-large baryon density, $\dent \ll
1$, both because it is the relevant regime for the RHIC and LHC
experiments and because it allows to derive an analytic solution.
Exploring the $\tilde\delta\sim 1$ regime requires a numerical
analysis, that we plan to provide in the near future.

The main results and the outline of the paper are as follows. In
section  \ref{sec:ansatz} we present the action and the ansatz for
the D3-D7 setup at finite baryon density. A set of second order
differential equations is given in terms of the functions of the
radial variable appearing in the ansatz. In section
\ref{sec:solution} we solve the equations analytically, in a
perturbative expansion in $\epshne$ and $\tilde\delta$ up to second
order in both parameters. In section \ref{sec:physics} we perform
the study of the thermodynamics of the system in the canonical and
grand-canonical ensembles, checking the (non-trivial) closure of the
various thermodynamic relations. We then explore the effects of the
baryon number density on the energy loss of probes through the
plasma, in particular on the jet quenching parameter. While the
overall effect of flavors is to enhance the jet quenching
\cite{D3D7QGP}, the effect of finite baryon density depends on the
specific choice of comparison scheme of different theories. We
finally provide some considerations on certain ``optical" properties
of the plasma, thinking about the possible gauging of the global
$U(1)$. Section \ref{sectionconclusions} contains some concluding
remarks. We also provide an appendix with some details on the
ten-dimensional action, the equations of motion, the Bianchi
identities and their solutions.

The backgrounds we provide correspond to charged black holes in
(slightly deformed) $AdS$, the charge being dual to a finite baryon
density. The regime of validity is completely specified and the
solution is totally reliable in that regime -- there are no
uncontrolled approximations. It is the first solution of this kind
in the literature and thus it is suitable for the study of a number
of physical effects of the baryon density. We hope to explore
further the physics of this system in the future.

\section{Ansatz  and effective Lagrangian}
\label{sec:ansatz} \setcounter{equation}{0}

The field theories we focus on are realized on the 4d intersection
of $N_c$ ``color" D3 and $N_f$ homogeneously smeared ``flavor"
D7-branes. The D3-branes are placed at the tip of a Calabi-Yau (CY)
cone over a Sasaki-Einstein manifold $X_5$, the latter being a
$U(1)$ fiber bundle over a four dimensional K\"ahler-Einstein (KE)
base. The ambient spacetime, a product of 4d Minkowski and the CY
cone, will be deformed by the backreaction of both kind of branes
which respectively source a (self dual) $F_5$ and a $F_1$ RR field.
As a result the 10d metric will be in the form of a warped product
and there will be a running dilaton. Moreover, the backreaction of
the D7-branes will induce a squashing between the KE base of the
Sasaki-Einstein manifold and the fibration \cite{Benini:2006hh}.

Finite temperature is realized by placing a black hole in the center
of the background \cite{D3D7QGP}. The D7-branes extend along the
radial direction up to the black hole horizon. Their embedding is
described by a constant profile, implementing massless flavor fields
in the dual gauge theories. In this work we are interested in
switching on a chemical potential for the $U(1)_B$ baryon symmetry.
The dual picture involves a non-vanishing profile for the temporal
component $A_t$ of the worldvolume gauge field on the D7-branes
\cite{hep-th/0611099}.
Through the Chern-Simons coupling, this field can source $F_3$ and $H_3$ form fields.

All in all we will be dealing with a general type IIB action given,
in Einstein frame, by
\beqa
S&=&\frac{1}{2\kappa_{10}^2}\Bigg[\int
d^{10}x\,\sqrt{-g} \left(
R - \frac{1}{2}(\d \Phi)^2 - \frac{e^{-\Phi} }{2} H_3^2 - \frac{ e^{2\Phi}}{2} F_{1}^2 - \frac{e^{\Phi} }{2}F_{3}^2 - \frac{1}{4}F_{5}^2\right) \nonumber \\
&&\qquad -\int C_4\wedge H_3\wedge F_{3}\Bigg] + S_{fl}\, ,
\label{coloract}
\eeqa
where
\be
S_{fl} = - T_7 \sum^{N_f}\int_{D7}
d^8\chi \,e^\Phi\sqrt{-\det(\hat g + e^{-\Phi/2}{\cal F})} + \mu_7
\sum^{N_f}\int_{D7}\hat C_q \wedge \left(e^{-\cal F}\right)_{8-q} \,
\label{flact}
\ee
is the contribution of the flavor D7-branes. The
gravitational constant and D7-brane tension and charge are, in terms
of string parameters
\be \frac{1}{2\kappa_{10}^2} = \frac{T_7}{g_s}
= \frac{\mu_7}{g_s} = \frac{1}{(2\pi)^7g_s^2 \alpha'^4}\,\,.
\label{kappa10} \ee

The smearing procedure \cite{Bigazzi:2005md,cnp} amounts to a
replacement \be \label{smear} \sum^{N_f} \int_{D7} X_8 \to
\int_{M_{10}} X_8 \wedge \Omega_2 \, , \ee for any form $X_8$
defined on the brane worldvolume.\footnote{The smearing of the DBI
part of the flavor branes is described at length in
\cite{Benini:2006hh}.} Here $\Omega_2$ is a form orthogonal to the
individual location of the D7-branes. For an arbitrary
Sasaki-Einstein space $X_5$, it is proportional to the K\"ahler form
$J_{KE}$ of the K\"ahler-Einstein 4d basis \cite{Benini:2006hh}
\be
g_s \Omega_2 = -2 Q_f J_{KE} \, .
\ee
For massless flavors, $Q_f$ is
a constant encoding the density of D7-branes in the relative
quotient space $X_5/X_3$ with $X_3$ the subspace wrapped by each of
the branes
\be Q_f = \frac{{\rm Vol}(X_3)  g_s N_f}{4\,{\rm
Vol}(X_5) } \, .
\label{Qfconstant}
\ee
The equations of motion and
Bianchi identities that follow from the action (\ref{coloract}),
(\ref{flact}) are given in appendix \ref{app:solut}.

Concerning the metric, we will consider the following ansatz, which
includes a family of generalized squashed Sasaki-Einstein manifolds
\be ds_{10}^2 = h^{-1/2}[ -b\, dt^2 + dx^i dx_i]  + h^{1/2}[ bS^8
F^2  d\sigma^2 +S^2 ds^2_{KE} + F^2(d\tau + A_{KE})^2]
\label{genansatz} \,, \ee where the  K\"ahler two-form of the four
dimensional base is given in terms of the connection one-form as
$J_{KE}=dA_{KE}/2$. The ansatz (\ref{genansatz}) contains two
squashing functions $F(\sigma)$ and $S(\sigma)$ (with dimension of
length), whose quotient $F/S$ parameterizes the effect of the flavor
backreaction. The dimensionless functions $h(\sigma)$ and
$b(\sigma)$ account for the warping and the blackening of the
spacetime, respectively. Thus, in particular, an ansatz with $b=1$
is appropriate for the zero-temperature, uncharged solution. We have
used the invariance under diffeomorphisms to choose a convenient
holographic radial direction, $\sigma$ (with dimension of
length$^{-4}$), and as we will see, $\sigma\to-\infty\, (0)$ in the
IR (UV). Given the smearing procedure (\ref{smear}) all the
functions in our ansatz depend only on the radial variable.

The finite baryon density is dual to a
nontrivial worldvolume $U(1)$ gauge field,
\be
{\cal F} = 2\pi \alpha' A_t'(\sigma) \, d\sigma\wedge dt\, .
\ee
A consistent ansatz for the other fields is
\bea
&& \Phi=\Phi(\sigma)\, , \qquad B_2=0 \, , \qquad F_1 = Q_f(d\tau + A_{KE})\, , \qquad F_5 = Q_c(1 + *){\mathcal{V}}(X_5) ~,~~~~~~~~~~~~~~\,  \label{ansatz} \\
\rule{0mm}{9mm}
&& F_3 = F_{123} dx^1\wedge dx^2 \wedge dx^3 - \frac{J' e^{-\Phi}}{S^4 F^2}\,  dt\wedge \Omega_2 + 8 Q_f J e^{-\Phi} b F^2\,  dt\wedge d\sigma\wedge (d\tau + A_{KE})\ .
\eea
In these expressions, ${\mathcal{V}}(X_5)$ is the volume form of $X_5$, $Q_c$ is proportional to the number of colors
\be
Q_c = \frac{(2\pi)^4 g_s \alpha'^2 N_c}{{\rm Vol}(X_5)} \, , \label{flavdens}
\ee
$F_{123}$ is a constant (of dimension length$^{-1}$) which we will
show to be related to the baryon density, whereas $J=J(\sigma)$ is a
function (of dimension length$^{3}$) that describes the effects of
the backreaction;\footnote{Obviously, $J(\sigma)\neq J_{KE}$.} its
contribution is dictated by the $C_6$ potential \be C_6 =
J(\sigma)\,  dx^1\wedge dx^2\wedge dx^3\wedge (d\tau + A_{KE})
\wedge \Omega_2 \, , \ee which is the natural D5 charge sourced on
the world-volume of the D7-branes by the gauge field through the
last term in (\ref{flact}).\footnote{Notice that, with the smearing,
$\displaystyle \sum_{N_f}\int_{{\cal M}_8}\hat C_6 \wedge {\cal F}
\to \int_{{\cal M}_{10}}C_6 \wedge {\cal F}\wedge \Omega_2$.} In
\cite{matsuuracfl} the system without backreacting flavors was
studied, and an ansatz for $F_3$ was used that only contained the
piece proportional to $F_{123}$. However, for our equations of
motion to be consistent, we need the presence of the other
components; thus, we see that $J(\sigma)$ naturally contains the
effects of the backreaction of the flavors.

Inserting the whole ansatz into the 10d equations of motion and
Bianchi identities one finally arrives at a system of equations
which the reader can find in formulas
(\ref{gaugefieldeom})--(\ref{eomJ}).
It is possible to describe the whole system in terms of an effective
one-dimensional action from which the equations of motion can be
derived \be S =\frac{{\mathrm{Vol}}(X_5)V_{1,3}}{2\kappa_{10}^2}
\int  L_{1D}  d \sigma\, , \ee where $V_{1,3}$ denotes the
(infinite) integral over the Minkowski coordinates, and
\beqa\label{onedimlag}
L_{1D} &=&  - \frac12 (\log' h)^2  +12(\log'S)^2 + 8 \log'F \log'S - \frac12 \Phi'^2 \nonumber\\
&& \left.
+\frac{\log' b}{2}\left( \rule{0mm}{3mm}\log' h+ 8\log'S + 2 \log'F  \right)-4 Q_f^2\frac{ J'^2}{F^2 S^4} \right. \label{1Dlag}\\
&&\left.- \frac{b Q_c^2}{2h^2} - 4 b F^4 S^4 + 24 b F^2 S^6 - \frac12 F_{123}^2 e^{\Phi} b  h^2 F^2  S^8 - \frac12Q_f^2 e^{2\Phi} b  S^8 \right. \nonumber \\
 && - 4 e^{\Phi/2} F Q_f S^2 \sqrt{-(2\pi\alpha'A_t')^2 + e^{\Phi}  b^2 F^2 S^8}
 - 32 Q_f^2e^{-\Phi} b F^2 J^2  - 8Q_f^2(2\pi \alpha' A_t') J \nonumber \, .
\eeqa The constraint equation (\ref{apconstraint}) is the zero
energy condition $H=0$ for the Hamiltonian \be
~~~~~~~~~~~~~~~~~~~~~~~~~~~~~H = -L_{1D} + \sum_i \psi_i' \frac{d
L_{1D}}{d \psi_i'}\, , ~~~~~~~~~~~~~~\psi_i= \{ b,h,F,S,\Phi,A_t,J \}
\, . \ee Since the gauge field $A_t$ enters only through its
derivative it leads to a ``constant of motion". In principle this is
a new free parameter which is related to the charge density. However
the equations of motion link this to the value of $F_{123}$ in the
ansatz for $F_3$ in (\ref{ansatz}) as we now show. Let us fix this
constant of motion as follows \be\label{densitydef} \frac{\partial
L_{1D}}{\partial A_t'}  \equiv 2\pi \alpha' Q_c F_{123} \,
.\label{conservat} \ee Solving for $A_t'$ it gives \be 2\pi\alpha'
A_t' =  \frac{(Q_c F_{123}+8  Q_f^2 J)bFS^4}{ \sqrt{16 Q_f^2 F^2
S^4+ e^{-\Phi}(Q_c F_{123}+8  Q_f^2 J)^2     }}
\label{gaugefieldeom2} \, . \ee Exactly the same expression is
obtained from the equation of motion for the form field $H_3$ (see
eq. (\ref{gaugefieldeom})). Thus, by enforcing  the integration
constant as in (\ref{conservat}) for consistency,  we are putting
the system partially on shell.  On the other hand, this obscures the
analysis when it comes to computing the thermodynamical potentials holographically,
since it means that the canonical momentum conjugate to $A_t'$ was
already present in the original Lagrangian. We will comment on this
later on.

It is natural to use  equation (\ref{gaugefieldeom2}) to eliminate
$A_t'$ in favor of $F_{123}$ and this is usually done in one of two
ways:  obtaining the equations of motion from \eqref{onedimlag}
and then imposing eq. \eqref{gaugefieldeom2}, or else,  performing a
Legendre transformation to the Lagrangian \be \tilde L_{1D}=
L_{1D}-\frac{\delta L_{1D}}{\delta A_t'}
A_t'\Bigg|_{A_t'=A_t'(F_{123})} \, ,\label{legtranLag1D} \ee and
then taking the Euler-Lagrange equations from the transformed
action. Either way the equations of motion coincide and are given by
(\ref{gaugefieldeom})--(\ref{eomJ}) in appendix A.

\section{The perturbative solution}
\label{sec:solution} \setcounter{equation}{0} In the uncharged case
$A_t'=F_{123}=J=0$, the following exact solutions for the functions
$b$ and $h$ are readily found \cite{D3D7QGP}: $b=e^{4r_0^4\sigma},\,
h=Q_c(1-e^{4r_0^4\sigma})/(4r_0^4)$, where $r_0$ is an integration
constant of dimension of length. The black hole horizon is at
$\sigma\rightarrow -\infty$ and the extremal limit is reached
sending $r_0\rightarrow 0$. In terms of a more standard radial
coordinate $r$, defined in such a way that $h=R^4/r^4$ with
$R^4=Q_c/4$, one gets $b=1-(r_0/r)^4$ precisely as for the
unflavored $AdS_5$ black hole. The horizon radius $r_h=r_0$ is related
to the temperature of the black hole. The whole solution in
\cite{D3D7QGP} also depends on the dimensionless combination
$\epsilon=Q_f e^{\Phi}$, which weighs the backreaction of the
D7-branes and, in fact, can be read as a flavor-loop counting
parameter in the dual field theory.

Now, $\epsilon$ runs as the dilaton and thus (as a common feature of
backreacted D3-D7 setups) it blows up at a finite scale $r_{LP}$
(corresponding to a UV Landau pole in the dual field theory),
rendering the supergravity approximation not reliable. Keeping
$\epsilon$   small requires restricting the validity range of our
solution up to an arbitrary cutoff $r^*\ll r_{LP}$, such that \be
\epsilon_* = Q_f e^{\Phi_*}  = \frac{{\mathrm{Vol}}(X_3)}{16\pi
{\mathrm{Vol}}(X_5)}\lambda_* \frac{N_f}{N_c}\ll 1 \label{epslambda}
\, , \ee where $\lambda_*=4\pi g_s e^{\Phi_*} N_c\gg1$, $N_c,
N_f\gg1$, $\Phi_*=\Phi(r_*)$ and we have used (\ref{flavdens}). In
the uncharged case \cite{D3D7QGP}, this condition was used to find
an analytic perturbative solution, up to order $\epsilon_*^2$, for
the remaining functions $S,F,\Phi$ appearing in the ansatz. The
related integration constants were fixed requiring regularity at the
horizon and matching with the $T=0$ solution \cite{Benini:2006hh} at
the UV cutoff $r_*$. The resulting functions thus contained the
dimensional (resp. dimensionless) parameters $r_h, r_*$ (resp.
$\epsilon_*$). The UV cutoff dependent terms resulted to be of the
form of both power-like and logarithmic corrections. Formally
sending the arbitrary cutoff scale $r_*$ to infinity, the first kind
of corrections drops out, while the second kind can be handled
taking into account that the function $\epsilon$ has a logarithmic
running (accounting for the breaking of conformal invariance induced
at the quantum level by the massless flavors) such that \be
\epsilon_h \equiv Q_f e^{\Phi(r_h)} = \epsilon_*\left(1+ \epsilon_*
\log\frac{r_h}{r_*}\right) + {\cal O}(\epsilon_*^3)\,.
\label{logrun} \ee This procedure allows to decouple the IR physics
from the UV one and to write down a set of solutions containing just
$r_h$ and $\epsilon_h$ as parameters.

In the present charged case, we are going to follow the very same
procedure. Here we have a further parameter to deal with: we will
call it $\tilde\delta$, and we will show that it is related to the
dimensionless combination of temperature and baryon chemical
potential (or charge density, depending on the thermodynamical
ensemble). We will then derive an analytical perturbative solution
taking both $\epsilon_*$ and $\tilde\delta$ to be much smaller than
one, deforming   the finite temperature flavor backreacted solution
obtained in \cite{D3D7QGP}.

As a first step, let us introduce a dimensional parameter $\den$
and consider the following redefinitions \be
  \quad F_{123} = \den  \frac{\sqrt{\epsilon_* Q_f}}{Q_c}\, ,~~~
  \quad  J(\sigma) = \den \frac{\epsilon_*^{3/2}}{Q_f^{3/2}} \tilde J(\sigma)  \label{scalingdelta}\, .
\ee The reason behind this choice will be clear in a moment.
Inserting these expressions into
(\ref{eomansatz})--(\ref{eomJ}) and rewriting the dilaton
as $\Phi(\sigma)=\Phi_* + \phi(\sigma)$, with $\phi(\sigma_*)=0$,
one readily arrives at the following system of equations \beqa
(\log b)'' &=& 4\, {\epsilon_*\den^2 \, }\, \frac{X}{Y} + 64\, \epsilon_*^2\den^2 \, e^{-\phi} \, b F^2  \tilde J^2 + 8\, \epsilon_*^2\den^2 \, e^{-\phi} \,  \frac{ \tilde J'^2}{F^2 S^4} + \epsilon_*^2\den^2 \,  \,Z\, , \nonumber\\
(\log h)'' &=& - Q_c^2 \frac{b}{h^2} + 2\, \epsilon_*\den^2 \,  \, \frac{X}{Y} + 32 \, \epsilon_*^2\den^2 \, e^{-\phi} \,b F^2 \tilde J^2 + 4\, \epsilon_*^2\den^2 \, e^{-\phi} \,\frac{\tilde J'^2}{F^2 S^4} + \epsilon_*^2\den^2 \,  \,\frac{3}{2}Z \, ,\nonumber\\
(\log S)'' &=& - 2 b F^4 S^4 + 6 b F^2 S^6 - \epsilon_* \, e^{3\phi/2}\frac{b^2 F^3 S^{10}}{Y} - 16\, \epsilon_*^2\den^2 \, e^{-\phi} \, b F^2 \tilde J^2  - \epsilon_*^2\den^2 \,  \,\frac{1}{4}Z \, ,\nonumber\\
(\log F)'' &=&~~ 4 b F^4 S^4 -\frac{1}{2}\epsilon_*^2 e^{2\phi}\, b S^8 - \epsilon_*\, \den^2 \, \, \frac{X}{Y} + 16\, \epsilon_*^2\den^2 \, e^{-\phi} \, b F^2 \tilde J^2 - 2 \, \epsilon_*^2\den^2 \, e^{-\phi} \,\frac{\tilde J'^2}{F^2 S^4}\nonumber \\
&& - \epsilon_*^2\den^2 \,  \,\frac{1}{4}Z\, , \nonumber\\
(\phi)'' &=&  \epsilon_*^2 e^{2\phi}\, b S^8 +2\, \epsilon_* \, e^{3\phi/2} \frac{b^2 F^3 S^{10}}{Y} + 2 \, \epsilon_*\,   e^{\phi/2} F S^2 Y -32 \, \epsilon_*^2\den^2 \, e^{-\phi} \,b F^2\tilde J^2  \nonumber\\
& & - 4 \, \epsilon_*^2\den^2 \, e^{-\phi} \,\frac{\tilde J'^2}{F^2 S^4}+ \epsilon_*^2\den^2 \,  \,\frac{1}{2}Z\, ,\nonumber\\
\left[ \frac{e^{-\phi}\tilde J'}{S^4 F^2} \right]' &=& \frac{( 1+8\epsilon_* \tilde J)bFS^4}{\sqrt{16 F^2 S^4+ \den^2e^{-\phi}(1+8\epsilon_* \tilde J)^2    }}
+8e^{-\phi}b  F^2 \tilde J \, ,
\label{eomscaling}
\eeqa
\medskip
with
\beqa
X &=&  \frac{(1+8 \epsilon_* \tilde J)^2   e^{\phi/2}b^2 F^3S^{10}}{ 16   F^2 S^4+\den^2 e^{-\phi}(1+8 \epsilon_* \tilde J)^2     }\, ,
 \\
Y &=& \sqrt{b^2 e^\phi F^2 S^8- \frac{\delta^2(1+8  \epsilon_* \tilde J)^2b^2F^2S^8}{  16   F^2 S^4+ \den^2 e^{-\phi}(1+8 \epsilon_* \tilde J)^2  } }\, ,
 \\
Z &=& \frac{e^{\phi} \, b h^2 F^2 S^8}{Q_c^2} \label{xyzscaling}\, .
\eeqa
The constraint equation (\ref{apconstraint}) reads
\bea
\nn 0 & = & - \frac12 \log' h \log' b + \frac12 (\log' h)^2 -
12(\log'S)^2 - 4\log'b \log'S\\
\nn & &  - \log'b \log'F - 8 \log'F \log'S + \frac12 \phi'^2\\
\nn & & - \frac{b Q_c^2}{2 h^2} - 4 b F^4 S^4 + 24 b F^2 S^6 -\epsilon_* \frac{4 e^{3\phi/2}b^2 F^3  S^{10}}{Y}  - \epsilon_*^2 \frac12 e^{2\phi} b  S^8\nonumber \\
& & +\epsilon_*^2 \delta^2\left(- 32   b e^{-\phi} F^2  \tilde J^2 +  \frac{4 e^{-\phi}   \tilde J'^2}{F^2 S^4}-\frac12 Z\right)\, .
\label{constra}
\eea
The system (\ref{eomscaling})--(\ref{constra}) allows for a
systematic expansion of all the functions in powers series of
$\epsilon_*$ and $\den^2$. This is essentially the main effect of the scaling relations (\ref{scalingdelta}).
Once all the functions have been solved for, the worldvolume gauge
field can be obtained from the following relation \be 2\pi\alpha'
A_t' =\den e^{\Phi_*/2} \frac{(1+8\epsilon_* \tilde
J)bFS^4}{\sqrt{16 F^2 S^4+ \den^2 e^{-\phi}(1+8\epsilon_* \tilde
J)^2 }} \label{gfieldprof} \, , \ee which is already first order in
$\den$. From this, we also deduce  (as previously announced) that
$J(\sigma)$ takes the effects of the flavor backreaction into
account.

In order to integrate the system (\ref{eomscaling})--(\ref{constra})
it is easier to switch to a radial coordinate $y=e^{4 r_0^4\sigma}$
with $r_0$ an arbitrary parameter of dimension of length. The
dimensionless parameter $\tilde\delta$ referred to as above is then
defined as \be \dent=\frac{\den}{4 r_0^3}\,, \ee where  the factor
$4$ is introduced in order to make it  precisely $\dent =\tilde d$
of ref. \cite{hep-th/0611099} (see also the equation
\eqref{chargedens} in this paper). We keep the greek symbol,
however, in order to stress that our parameter is going to be
perturbatively small.

Following the same procedure as in the uncharged case, we are now
able to provide analytic solutions to the equations above, in a
perturbative expansion in $\epsilon_*$ and $\tilde\delta$. We will skip here the
intermediate step where the solutions contain cutoff-dependent terms
(this can be found in appendix \ref{app:solut}  up to order
$\epsilon_*\,\dent^2$ in eqs. (\ref{solubst}) to (\ref{soluJst}))
and focus on their effective IR expressions. Introducing the IR
parameter $\epsilon_0=\epsilon(r_0)$ and the radial variable $r$
(defined again in such a way that the warp factor keeps the standard
$AdS$ form), they read
\beqa
h(r) &=& \frac{R^4}{r^4}\, ,  \label {hqb}\\
b(r)&=& \left(1 - \frac{r_0^4}{r^4}\right)
- \frac{\dent^2\epsilon_0}{2}\left( \left( 2-\frac{r_0^4}{r^4} \right) \left(\frac{r_0^2}{r^2}-\log\left[1+\frac{r_0^2}{r^2}\right]\right)\right)
\label{effeqb}
\\
&&
+\frac{\dent^2\epsilon_0^2}{12}\left( 17\frac{r_0^2}{r^2}-9\frac{r_0^4}{r^4}-\frac{5}{2}\frac{r_0^6}{r^6} - \frac{17}{2}\left(2-\frac{r_0^4}{r^4}\right)
\log(1+\frac{r_0^2}{r^2})\right)+...\, ,
 \nonumber
\\
S(r) &=& r \left[
1+ \frac{\epsilon_0}{24} + \epsilon_0^2\left(\frac{9}{1152} - \frac{1}{24}\log\frac{r_0}{r}\right)
\right.
\\
&&  + \frac{\epsilon_0\dent^2}{40}\left( 3 - 2\frac{r^2}{r_0^2} - 3\left(1-2\frac{r^4}{r_0^4}\right)\log\left[1+\frac{r_0^2}{r^2}\right]-\frac{1}{2}G(r)\right)
\nonumber\\
&& \left.
+ \frac{\epsilon_0^2\dent^2}{320} \left( -33 + 22\frac{r^2}{r_0^2} + 33\left(1- 2\frac{r^4}{r_0^4}\right)\log\left[1 + \frac{r_0^2}{r^2}\right] + \frac{11}{2}G(r)\right)+...\right]\, ,
\\
&&
\nonumber
 \\
F(r) &=&  r\left[  1 -  \frac{\epsilon_0}{24} + \epsilon_0^2 \left( \frac{17}{1152} + \frac{1}{24}\log \frac{r_0}{r}\right)   \right. \\
&& \quad +  \frac{\epsilon_0\dent^2}{40}\left( 3 -22\frac{r^2}{r_0^2} +5\frac{r_0^2}{r^2} -3 \left(1-2 \frac{r^4}{r_0^4} \right) \log\left[1 + \frac{r_0^2}{r^2}\right] + 2G(r)   \right)
\nonumber\\
&& \left. \quad +  \frac{\epsilon_0^2\dent^2}{192} \left(-21+ 154\frac{r^2}{r_0^2} -35 \frac{\rhne^2}{r^2} + 21 \left(1-2\frac{r^4}{\rhne^4}\right) \log\left[1+ \frac{\rhne^2}{r^2}\right]  - 14\, G(r)  \right) +... \right] \, ,\nonumber\\
\Phi(r) &=& \Phi_0 + \epsilon_0\log\frac{r}{\rhne}  - \frac{\epsilon_0^2}{48}\left(8 \left(1+3\log \frac{r}{\rhne}\right) \log\frac{\rhne}{r} - 3 \,\mathrm{Li}_2\left[1-\frac{\rhne^4}{r^4}\right]\right)
  \label{phisolrh}\\
     &&\quad + \frac{\epsilon_0^2\dent^2}{120}\left(  26\left(1- \frac{r^2}{\rhne^2}\right) -2\pi - 15\frac{\rhne^2}{r^2} +  \left(11 + 18\frac{r^4}{\rhne^4}\right) \log\left[1 + \frac{\rhne^2}{r^2}\right]-14\log 2 \right. \nonumber \\
&& \quad + G(r)+... \Bigr)\, ,\nonumber \\
     \tilde J(r) & = & - \frac{\rhne^3}{8} + ...\,,  \\
A_t(r) &=& \frac{r_0}{4\pi\alpha'}\tilde\delta
e^{\Phi_0/2}\left(1-\frac{\epsilon_0}{6}\right)\left(1-\frac{r_0^2}{r^2}\right)
+...\, ,\label{jsolrh} \eeqa where $G(r)=2\pi
\frac{\rhne^6}{r^6}\,{}_2F_{1}\left(\frac{3}{2},\frac{3}{2},1,1-\frac{\rhne^4}{r^4}\right)$
is an hypergeometric function and $Li_2(u)\equiv \sum_{n=1}^\infty
\frac{u^n}{n^2}$ is a polylogarithmic function. Notice that $J$
enters at order $\epsilon_0^2$ in the equations, hence only the
leading contribution in $\tilde J$ is relevant in the solution.

The above solution must be supplemented with a Jacobian factor for
the change of radial coordinate $  Y(r) =d\sigma/d r $ which will
show up in the coefficient of $dr^2$
 \beqa
Y(r) &=&   \frac{1}{r(r^4-r_0^4)} +
\frac{\dent^2}{4r^3(r^4-r_0^4)^2}
 \left[ \epsilon_0 \left( r_0^2 (2r^2-r_0^2)(r^2+r_0^2) - 2r^6\log\left(1 + \frac{r_0^2}{r^2}\right) \right)
 \right.
 \nonumber \\
 &&
 \left.
 +\, \frac{\epsilon_0^2 }{12}
 \left( - 7 r_0^6 + 19 r_0^4 r^2 - 34 r_0^2 r^4 + 34 r^6 \log\left( 1 + \frac{r_0^2}{r^2}\right) \right) \right]+...
 \eeqa

\section{Physical properties of the dual plasmas}
\label{sec:physics}
\setcounter{equation}{0}
\subsection{Thermodynamics}

In the previous section we have derived a solution that, in essence,
is a black hole dressed by a set of scalar (dilaton), vector
(Maxwell), as well as higher rank tensor fields. Now we are going to
extract the thermodynamical properties of the solution, providing,
in turn, a first non trivial validity check of the latter by
verifying the closure of the standard thermodynamical formulae. As
in \cite{D3D7QGP}, all
 quantities are obtained in power series of our perturbative expansion parameters and, therefore, the relevant
 thermodynamic relations can only be verified up to the relevant order.

To begin with, let us  stress that $\rhne$  does not coincide with
the horizon radius $\rhor$. This radius  is defined by $b(\rhor) = 0
+{\cal O}(\epsilon_0^3,\dent^4)$ and is perturbatively shifted by the
baryon density\footnote{In fact, there are two radii $r_\pm$ that solve $b(r_\pm)=0+{\cal
O}(\epsilon_0^3,\dent^4)$. This  is reminiscent of the case in the
Reissner-Nordstrom black hole. The value presented here corresponds
to the external radius $r_h=r_+$, i.e. the event horizon. Numerically we have checked that as $\tilde \delta$ increases
these two radii approach each other. However,  a potentially extremal black hole $r_-=r_+$ cannot be obtained  within the range
of validity of our solution, which is perturbative in $\tilde \delta$.
} \be
\rhor = \rhne \left( 1 + \frac{\epsilon_0 \dent^2}{8} (1  - \log 2)  -
\frac{\epsilon_0^2 \dent^2}{96}(11 - 17 \log 2) + ... \right)\, . \ee
Notice that $r_h > r_0$. Notice moreover that
\be
\epsilon_h\equiv\epsilon(r_h)=\epsilon_0 + {\cal O}(\epsilon_0^3)\,,\qquad e^{\Phi_h}\equiv e^{\Phi(r_{h})} = e^{\Phi_0} \left(1 + \frac{\epsilon_0^2}{8} \tilde\delta^2 (1-\log2)\right)\,,
\ee
so that we can trade $\epsilon_0$ for $\epsilon_h$, $\tilde\delta e^{\Phi_0}$ for $\tilde\delta e^{\Phi_h}$, and so on, in all of our expressions. In particular, looking at (\ref{jsolrh}), we get, to leading order in our expansion
\be
A_t(r) = \frac{r_h}{4\pi\alpha'}\tilde\delta e^{\Phi_h/2}\left(1-\frac{\epsilon_h}{6}\right)\left(1-\frac{r_h^2}{r^2}\right)\,,
\label{atrh}
\ee
which vanishes at the horizon as required to ensure IR regularity.

The temperature can be computed in the usual way giving \be T =
\frac{\rhne}{\pi R^2}\left(1- \frac{\epshne}{8} \left(1 +
\dent^2\right) - \frac{13}{384}\epshne^2\left(1 -\frac{2}{13}\dent^2
\right) + ... \right) \, . \label{tempr0} \ee The entropy density is
derived from the horizon's area \be\label{entropy} s =  \frac{1}{2}
\frac{\pi^5}{{\mathrm{Vol}}(X_5)} N_c^2T^3\left[1+
\frac{1}{2}\epshne(1 + \dent^2) + \frac{7}{24}\epshne^2(1 +
\dent^2)\right] \, . \ee As for other thermodynamical variables,
this expression, at first order in $\epsilon_h$ and
$\tilde\delta^2$, precisely reproduces (for the case of massless
flavors) the one found in the probe approximation in
\cite{hep-th/0611099}. The ${\cal O}(\epsilon_h^2)$ terms are instead
completely new.

Concerning the charge density, a proper value is given by the
integration constant in (\ref{conservat}), as its definition
coincides precisely with the electric field displacement. In terms of scaling
constants we have, to leading order \be \frac{d S}{d F_{t\sigma}}
=\frac{{\mathrm{Vol}}(X_5)}{2\kappa_{10}^2} \frac{\d L_{1D}}{\d
A_t'} = \frac{{\mathrm{Vol}}(X_5)}{ (2\pi)^7 g_s^2 \alpha'^4}2\pi
\alpha'  e^{\Phihne/2}Q_f\dent \, 4r_h^3\, , \ee and making use of
(\ref{Qfconstant}) as well as (\ref{tempr0}) this may be casted in
the form \be \quarkdens = \pi^{7/2}   \frac{N_c^2 }{
{\mathrm{Vol}}(X_5)^{1/2} }\, \frac{T^3}{ \sqrt{\lambda_h}}
\,\epshne \dent\, \left(1 + \frac{3}{8}\epshne\right)\, .
\label{chargedens} \ee
 Using (\ref{epslambda}) with $r_*\to r_h$ this can be written as
 \be
 \quarkdens = \frac{\pi^{5/2}}{16} N_f N_c \frac{{\mathrm{Vol}}(X_3)}{{\mathrm{Vol}}(X_5)^{3/2}} \sqrt{\lambda_h} T^3 \dent\, \left(1 + \frac{3}{8}\epshne\right)\, . \label{chargedens2}
 \ee
At leading order, and for the case ${\mathrm{Vol}}(S^3)=2\pi^2$ and
${\mathrm{Vol}}(S^5)=\pi^3$,  we exactly recover formula (A.11) in
\cite{hep-th/0611099}.\footnote{Beware that $\lambda_{ours}=4\pi g_s
e^{\Phi} N_c= 2\lambda_{theirs}$.} In the expressions above,
$\quarkdens$ is the quark density of the system, related to the
baryon density $\bardens$ by the number of colors $\quarkdens=N_c\,
\bardens$.

\subsection{Thermodynamical Potentials}

We proceed now to the calculation of the Helmholtz and Gibbs free
energies, $\cal F$ and $\Omega$ respectively.\footnote{The
difference of these quantities with respect to the forms defined in
section \ref{sec:ansatz} should be clear from the context.} These
can be either directly evaluated starting from the expression for
the entropy density and using the standard thermodynamical
relations, or they can be deduced holographically. In the latter
case they are identified with the (renormalized) on-shell boundary
action for the gravity background, evaluated in the corresponding
ensemble. Consistency of the solution requires that whatever method
is chosen the results are the same.\footnote{In the following we
will not report the details of the holographic calculations, but we
will just give a sketch of the needed ingredients. Needless to say,
we have verified that the results deduced from thermodynamical
relations agree with those found from holography.}

A quick look at the 1$D$ effective Lagrangian given in
(\ref{onedimlag}) reveals that on one hand $A_t$ is a cyclic
coordinate and, on the other hand, $F_{123}$ is a Lagrange multiplier.
Amusingly enough, the fact that the equations of motion for $H_3$
and $A_t$ are consistent with one another imposes (\ref{conservat}),
which is nothing but the statement that, up to constant factors,
$F_{123}$ and $A_t$ are canonically conjugate variables. Hence, as it stands, $L_{1D}$  contains both $A_t'$ and
$F_{123}$, hence velocities and momenta.
As  a consequence the associated
action corresponds to neither the canonical nor the
grand-canonical ensemble.
\subsubsection{Canonical ensemble}

The Legendre transformed Lagrangian $\tilde L_{1D}$ given in
\eqref{legtranLag1D} is the natural one to describe the system in
the canonical ensemble, since it is fully expressed  in terms  of
the baryon density  parameter $F_{123}$. Therefore, evaluating the
associated action on-shell  we should obtain the Helmholtz free
energy, $\cal F$. As it is well-known, the action has to be
supplemented with the standard Gibbons-Hawking term to deal with a
well-posed variational problem. Even with this addition, the
evaluation presents divergences which we deal with by subtracting
the same quantity evaluated on the Euclidean solution at the same
temperature but without a horizon and also with no chemical
potential.  This procedure is explained in appendix B of
\cite{D3D7QGP}, where we refer the reader  for details. We obtain
the Helmholtz free energy density \be f=\frac{\cal F}{V_{3}} =
-\frac{1}{8} \frac{\pi^5 }{{\mathrm{Vol}}(X_5)} N_c^2 T^4 \left[ 1+
\frac{1}{2} \epshne\left( 1-2\dent^2 \right) + \frac{1}{6}\epshne^2
\left( 1-\frac72 \dent^2 \right) \right] \, , \label{freeendens} \ee
which, consistently, satisfies the thermodynamic relation $-\partial
f/\partial T = s$. To check this relation it is very important to
consider the dependence of $\epshne$ and $\dent$ on $T$.

The logarithmic running of $\epsilon =Q_f e^{\Phi(r)}$ (see also eq. (\ref{logrun})) and the map between the horizon radius and $T$ give \cite{D3D7QGP}
\be\label{eq:epsdependence} \frac{\d \epshne}{\d T} =
\frac{\epshne^2}{T} + {\mathcal{O}}(\epshne^3,\dent^2) \,
.\label{evolepsh} \ee Moreover, for $\lambda_h(T)= 4\pi g_s e^{\Phihne(T)}
N_c$ one gets by the same token
\be\label{eq:lamdependence} \frac{\d \lambda_h}{\d T}
=\epshne \frac{\lambda_h}{T} + {\mathcal{O}}(\epshne^2,\dent^2) \,
.\label{evolepsh} \ee In the canonical ensemble we must keep the
physical (dimensional) charge density  invariant. From equation
(\ref{chargedens2}) we see that the dependence of $\dent$ on $T$ at
fixed $\bardens $ comes from solving  as \be \dent(T) =
\frac{\bardens \, \alpha}{\sqrt{\lambda_h}T^{3}} \left(1 + {\mathcal
O}(\epshne) +...\right)\, , \ee with $\alpha$ a  $T$-independent
constant. Using (\ref{eq:lamdependence}) we obtain \be
 \left( \frac{d \, \dent(T)}{d T}\right)_{\bardens} = - \frac{\dent}{T} \left( 3 +\frac{\epshne}{2}  + {\mathcal O}(\epshne^2) \right)\, . \label{dentc}
\ee
Using (\ref{eq:epsdependence}) and (\ref{dentc}) it is straightforward to check that $-\partial f/\partial T = s$ with $f$ and $s$ given in
(\ref{freeendens}) and (\ref{entropy}) respectively. This is a strong proof of consistency.

Next we can evaluate the ADM energy density of the plasma, again
subtracting the contribution from the zero temperature and baryon
density setup to get rid of the divergences. Following appendix B in
\cite{D3D7QGP} the final result is \be\label{ADMresult} \varepsilon
= \frac{E_{ADM}}{V_{3}} = \frac{3}{8} \frac{\pi^5
}{{\mathrm{Vol}}(X_5)} N_c^2 T^4  \left[ 1+ \frac{1}{2} \epshne
\left( 1+2\dent^2 \right) + \frac{1}{3} \epshne^2\left( 1+\frac74
\dent^2 \right)  \right] \, , \ee which satisfies the relation
$\varepsilon = f+s T$. From here, and taking again into account
(\ref{eq:epsdependence}) and (\ref{dentc}), we obtain  the  heat
capacity at fixed baryon number density \be c_{V,\quarkdens} =
\left(\frac{\partial \varepsilon}{\partial T}\right)_{V, \bardens} =
\frac{3}{2} \frac{\pi^5 }{{\mathrm{Vol}}(X_5)} N_c^2 T^3   \left[ 1+
\frac{1}{2} \epshne \left( 1-\dent^2 \right) + \frac{1}{24}
\epshne^2\left( 11-7\dent^2 \right)  \right] \, . \ee

\subsubsection{Grand-canonical ensemble}

We would like to obtain the thermodynamic quantities corresponding
to the grand-canonical ensemble. It would be tempting to think that
the correct Lagrangian density to use here is the original one
$L_{1D}$ in (\ref{1Dlag}). However, this is not the case. As
mentioned before, the fact that we have identified the canonical
momentum conjugate to $A_t'$ with  $F_{123}$ interferes with the
Legendre transform, since we  see that this parameter already
appears in $L_{1D}$. Were it not for this fact, the following
inverse Legendre transform \be \tilde{\tilde L}_{1D}= \tilde
L_{1D}-\frac{\d \tilde L_{1D}}{\d F_{123}}  F_{123}
\Bigg|_{F_{123}(A_t)} \, ,\label{invlegtr} \ee where  $F_{123}(A_t')
$ comes from solving \be A_t' = \frac{\d \tilde L_{1D}}{\d
F_{123}}\, , \ee would bring us back to the original lagrangian
$\tilde{\tilde  L}_{1D} = L_{1D}$. However, notice the presence of
the term with $F_{123}^2$ in (\ref{1Dlag}), which upon
(\ref{invlegtr}) will  change sign. Therefore, the relevant
Lagrangian for the computation of the Gibbs free energy is given in
(\ref{1Dlag}) with a sign flip in the term with $F_{123}^2$. Doing
that, and following the same steps as for $f$, we arrive at the
result \be \omega=\frac{\Omega}{V_{3}} = -p= - \frac{1}{8}
\frac{\pi^5 }{{\mathrm{Vol}}(X_5)} N_c^2 T^4 \left[ 1+ \frac{1}{2}
\epshne\left( 1+2\dent^2 \right) + \frac{1}{6} \epshne^2 \left(
1+\frac72 \dent^2 \right)  \right]  \, , \ee where $p$ is the
pressure. Now we can compare this expression with the one for the energy
density given in (\ref{ADMresult}). The interaction energy, given by
\be \frac{\varepsilon - 3p}{T^4} = \frac{\pi^2 N_c^2}{16
{\mathrm{Vol}}(X_5)}\epshne^2\, , \ee is just the same as in the
uncharged case. Hence, charge density of massless flavors does not
contribute to the breaking of conformal invariance. This is not
unexpected in field theory and, in a dual gravity picture, it is the
same as what happens in the case of the Reissner-Nordstr\"om AdS
black hole, where the presence of a net charge does not spoil the
relation $\varepsilon = 3p$.

We now would like to check the thermodynamic relations in the
present ensemble. Notice that the interpretation of $\dent$ has
changed and, henceforth, also its behavior with the temperature. The
reason is that   the physical parameter to fix in the
grand-canonical ensemble is the dimensional  chemical potential,
$\mu$. The difference among the two thermodynamical potentials
gives \be \mu\, \quarkdens = f-\omega =  \frac{\pi^5
}{4{\mathrm{Vol}}(X_5)} N_c^2 T^4 \epshne\dent^2 \left(
1+\frac{7}{12}\epshne \right)  \, . \ee Using \eqref{chargedens} we
can extract the chemical potential\footnote{Using the particular
value of $\mathrm{Vol}(X_5)=\pi^3$ we get the same expression at
leading order in $\epshne$ as in eq. (A.10) in \cite{hep-th/0611099}
after identifying $\tilde\mu = \tilde\delta/2$.}

\be \mu = \frac{\pi^{3/2}}{4{\mathrm{Vol}}(X_5)^{1/2}}\,
\sqrt{\lambda_h}\; T \, \dent  \left( 1+\frac{5}{24}\epshne
\right)\, . \label{chempotef} \ee

Now, the value of the world-volume gauge field in the UV (i.e. at $r\gg r_h$) is (see eq. (\ref{atrh}))
\be
A_{t,UV}=\frac14 \sqrt{\frac{\lambda_h}{\pi
{\mathrm{Vol}}(X_5)}}\pi^2\ T\ \dent \left(1 -\frac{\epshne}{24}
\right)\, . \ee Comparing the variation of the grand potential with
respect to $A_{t,UV}$ \be \frac{\delta \omega}{\delta A_{t,UV}}=-
\pi^{7/2}\frac{N_c^2 }{ {\mathrm{Vol}}(X_5)^{1/2} }\, \frac{T^3}{
\sqrt{\lambda_h}} \,\epshne \dent\, \left(1 +
\frac{5}{8}\epshne\right)\, , \ee with the relation \be \frac{\delta
\omega}{\delta \mu}=-n_b\, , \ee and taking into account formula
(\ref{chargedens}), we get the connection between the UV value of
the gauge field and the chemical potential \be A_{t,UV}=\mu\,
\left(1 - \frac{1}{4}\epshne\right)\, . \ee Note that this differs
at subleading order from the result $A_{t,UV} = \mu$ as obtained in
\cite{hep-th/0611099}; this is thus an effect of the backreaction of
the flavors.

From (\ref{chempotef}) we see that for fixed  $\mu$, $\dent$
acquires now  a dependence on $T$ like \be \tilde\delta(T) =
\frac{4}{\pi^2}\sqrt{\frac{\pi
{\mathrm{Vol}}(X_5)}{\lambda_h(T)}}\frac{\mu}{T} \left( 1 -
\frac{5}{24}\epshne \right)\, .\label{dentoft} \ee Using
(\ref{eq:epsdependence}) again and (\ref{dentoft}) we now get \be
 \left( \frac{d \, \dent}{d T}\right)_\mu = - \frac{\dent}{T} \left( 1 +\frac{\epshne}{2}  + {\cal O}(\epshne^2) \right)\, . \label{dentgc}
\ee With this scaling (and equation (\ref{eq:epsdependence})) it is
easy to check that the thermodynamic relation $-\partial
\omega/\partial T = s$ holds at the required order. Using again
(\ref{dentgc}) we can obtain readily the  heat capacity at fixed
chemical potential \be c_{V,\mu} = \left(\frac{\partial
\varepsilon}{\partial T}\right)_{V,\mu} = \frac{3}{2} \frac{\pi^5
}{{\mathrm{Vol}}(X_5)} N_c^2 T^3   \left[ 1+ \frac{1}{2} \epshne
\left( 1+\dent^2 \right) + \frac{1}{24}\epshne^2 \left( 11+7\dent^2
\right) \right] \, , \ee from which we can extract the speed of
sound \be c_s^2 = \frac{s}{c_{V,\mu}} = \frac{1}{3} \left(
1-\frac{1}{6}\epshne^2 \right) \, . \ee The parameter
$\delta_c=1-3c_s^2$ is related to the breaking of conformality as
before. As it happens for the interaction energy, it does not
receive corrections from the presence of finite baryon density.
Using the arguments in \cite{D3D7QGP}, this suggests that the bulk
viscosity is not affected by the presence of a finite baryon density
on the system.\footnote{There is some mismatch between our results
for the speed of sound and the heat capacities and corresponding
results found in the literature in the probe approximation. The
precise closure of the thermodynamical relations and the consistent
independence of the conformality breaking effects from the baryon
chemical potential in the massless flavored case, let us be
confident of the correctness of our results.}

\subsubsection*{Susceptibilities}

The ``quark" susceptibility\footnote{As usual, ``quark" is an abuse
of language for ``fundamental matter".} is \be \chi
=-\frac{\partial^2 \omega}{\partial \mu^2}=\frac{\partial
\quarkdens}{\partial \mu}=\frac{\pi
{\mathrm{Vol}}(X_3)}{4{\mathrm{Vol}}(X_5)}\ N_fN_c\ T^2
\left(1+\frac16\epshne \right)\, . \ee At leading order this matches
with the result obtained in the probe approximation in \cite{mst}
for the flavored ${\cal N}=4$ SYM case (where
$\mathrm{Vol}(X_3)=2\pi^2\,,\mathrm{Vol}(X_5)=\pi^3$). The other
three susceptibilities are \be -\frac{\partial^2 \omega}{\partial
T^2}=\frac{\partial s}{\partial
T}=\frac{3\pi^5}{2{\mathrm{Vol}}(X_5)}\ N_c^2\ T^2
\left(1+\frac12\epshne + \frac{11}{24}\epshne^2 \right)+\frac{\pi
{\mathrm{Vol}}(X_3)}{4{\mathrm{Vol}}(X_5)}\ N_fN_c\ \mu^2\
\left(1+\frac16\epshne \right)\, , \ee and \be -\frac{\partial^2
\omega}{\partial \mu \partial T}=-\frac{\partial^2 \omega}{\partial
T \partial \mu}=\frac{\pi
{\mathrm{Vol}}(X_3)}{2{\mathrm{Vol}}(X_5)}\ N_fN_c\ \mu\ T
\left(1+\frac16\epshne \right)\, . \ee The determinant of the
susceptibility matrix equals $\chi C$ where \be
C=\frac{3\pi^5}{2{\mathrm{Vol}}(X_5)}\ N_c^2\ T^2
\left(1+\frac12\epshne + \frac{11}{24}\epshne^2 \right)-\frac{3\pi
{\mathrm{Vol}}(X_3)}{4{\mathrm{Vol}}(X_5)}\ N_fN_c\ \mu^2\
\left(1+\frac16\epshne \right)\, . \ee The second term is
parameterically smaller than the first one, thus the theory is
thermodynamically stable.

\subsection{Probe parton energy loss}

In order to estimate how the finite charge density or chemical
potential influences the energy loss of an energetic probe parton
traveling through the plasma, we will make use of the results in
\cite{Baier,liu}. In this approach, the parton looses energy through
bremsstrahlung due to its interactions with the strongly coupled
medium. The amount of energy loss, which is ultimately the cause of
the jet quenching in the strongly coupled plasma, is encoded in a
transport coefficient termed $\hat q$, the ``jet-quenching
parameter". The latter can be derived, using the eikonal
approximation at high energy, as the coefficient of $L^2$ in an
almost light-like Wilson loop with dimensions $L^{-}\gg L$. The
Wilson loop is easily calculated in string theory.

In \cite{aredmas} a formula to extract the jet quenching parameter
from a general dual gravity background was derived.\footnote{We took
into account a different factor of $\sqrt2$ between the definitions
in \cite{liu} and \cite{aredmas}.} On our solution, rewritten in the variable $y=\exp(4 \rhne^4 \sigma)$, the formula gives \be
\hat q^{-1}= \pi\, \alpha' \int_{0}^{1} e^{-\frac{\Phi}{2}}
\frac{\sqrt{g_{yy}}}{g_{xx}\sqrt{g_{xx}+g_{tt}}}dy =\pi \int_{0}^{1}
\frac{1}{4 \rhne^4 y e^{\frac{\Phi}{2}}} \sqrt{\frac{b}{1-b}}h F S^4
dy \, , \ee where in particular $g_{xx}$ is the metric coefficient
in the spatial direction involved in the trajectory of the parton.
For our purposes, it will be enough to analyze the result up to
order $\epshne \dent^2$, which reads \be \hat
q=\frac{\pi^3\sqrt{\lambda_h}\Gamma(\frac34)}{\sqrt{{\mathrm{Vol}}(X_5)}\,\Gamma(\frac54)}T^3
\left[1 + \frac{2+\pi}{8} \epshne \Bigl(1+  c \, \dent^2\Bigr)
\right]\, , \label{jetq} \ee with a positive constant $c = 0.867565
$.\footnote{The constant $c$ is related to the result of the
integral \be
 \int_{0}^{1}\frac{-(8-3\sqrt{1-y}+2y)(1-y)+\sqrt{1-y}(3+2y)(2 \rm{arctanh}{[\sqrt{1-y}]}+\log{y})}{(1-y)^{9/4}\sqrt{y}}dy \, .\nonumber
\ee
}

\subsubsection{Comparison schemes}
\label{sectioncompare}

Formula (\ref{jetq}), without any prescription to compare different
theories, would imply that a finite charge density (or chemical
potential) increases the jet quenching. This conclusion would depend
on considering the jet quenching parameter in theories with
different numbers of degrees of freedom. From a phenomenological
perspective, it is more useful to compare theories keeping the
number of degrees of freedom fixed. In order to fully appreciate the
effects of flavors at finite charge density, we will compare our
flavored theory with the unflavored ${\cal N}=4$ SYM.

For an estimate of the number of degrees of freedom, the most used
observables are the entropy density (\ref{entropy}), and the energy
density (\ref{ADMresult}). In large $N_c$ theories, it makes sense
to keep one of these two quantities fixed by either fixing the
number of colors $N_c$ and changing the temperature $T$, or the
other way around. First, let us consider the fixed $N_c$, varying
$T$ comparison scheme, used for example in \cite{Gubser:2006qh}. For
the sake of definitiveness, we also keep the energy density fixed;
the qualitative result which will follow is unchanged if we fix the
entropy density instead. We get \be\label{compareT}
T_f=T_u\left[1-\frac{\epshne}{8}\Bigl(1+2\dent^2\Bigr) \right] \, ,
\ee where the subindex refers to either the $f$lavored or the
$u$nflavored ${\cal N}=4$ theory. Plugging this result in formula
(\ref{jetq}) we read \be\label{ratiojet} \frac{\hat q_f}{\hat q_u}
\sim \sqrt{\frac{\lambda_f}{\lambda_u}}  \left[ 1+
\frac{\pi-1}{8}\epshne\left( 1 - 0.719\;  \dent^2\right)\right]  \,
, \ee where both 't Hooft couplings $\lambda_f, \lambda_u$ are
evaluated at the horizon. Since $N_c$ is fixed,\footnote{Remember
that $\lambda \sim \alpha_s N_c$.} if we keep the coupling
$\alpha_s$ fixed, we obtain the result that while the jet quenching
parameter is enhanced by the addition of flavors \cite{D3D7QGP}, the
finite charge density (or chemical potential) actually reduces the
enhancement.\footnote{Remember that the term in $\epshne \dent^2$ is
by definition a perturbation of the one in $\epshne$.}

This qualitative result is unchanged if, following strictly
\cite{Gubser:2006qh}, we also allow for the variation of $\alpha_s$.
The latter can be adjusted in such a way that the force between two
external quarks at the screening length $L_{c,u}$ of the unflavored
plasma, $\alpha_{qq}=3 L_{c,u}^2 V'(L_{c,u})/4$, is kept fixed. The
potential $V(L)$ is calculated numerically by standard formulas,
reported for example in \cite{Kinar:1998vq}. We calculate
$\alpha_{qq}$ for different values of $N_f$ (equivalently,
$\epshne$) and $\dent$. At zero charge density, as we increase
$N_f$, $\alpha_{qq}$ increases, so in order to keep it fixed the
coupling $\lambda_f$ must be decreased.\footnote{This corrects a
statement in footnote 20 of \cite{D3D7QGP}.} But the effect is
small, so as anticipated we get  in (\ref{ratiojet}) that the jet
quenching parameter is enhanced by the addition of flavors. Along
the same lines, if we switch on the charge density, we find that
this enhancement is reduced.

We can now analyze what happens if we
use the alternative comparison scheme introduced in \cite{D3D7QGP},
where the temperature $T$ is kept fixed and the number of degrees of
freedom is kept constant by varying the number of colors $N_c$. By
fixing the energy density (\ref{ADMresult}) we get \be
N_{c,f}=N_{c,u}\left[1-\frac{1}{4}\epshne \Bigl(1+2\dent^2\Bigr)
\right]\, . \ee Keeping fixed also the coupling constant $\alpha_s$
for simplicity, we obtain \be \frac{\hat q_f}{\hat q_u} \sim 1+
\frac{\pi+1}{8}\epshne\left(1 + 0.594\; \dent^2\right) \, . \ee
Thus, while again the overall effect of the flavors is to enhance
the jet quenching, in this case the charge density actually
increases the enhancement. The result is qualitatively the same if
we keep fixed the entropy density (\ref{entropy}) instead of the
energy density.

To summarize, while in all the cases the effect of fundamental
flavors is to enhance the jet quenching \cite{D3D7QGP}, the net
effect of a finite charge density (or chemical potential) actually
depends on how we compare different theories. Namely, we found that
the charge density reduces (resp. increases) the enhancement of the
jet quenching due to flavors in the varying $T$, fixed $N_c$ (resp.
fixed $T$, varying $N_c$) comparison scheme. We currently have no
intuitive explanation for this behavior.

\subsection{Remarks on ``optical" properties}

The solution presented in this paper is dual, in the hydrodynamic
limit, to a charged relativistic fluid. Recently, it has been
observed in \cite{ragazzi1,ragazzi2} that, generically, this kind of fluids has
interesting uncommon optical properties, such as negative refractive
index, and exhibits exotic phenomena, like the propagation of additional light waves in certain
frequency regimes. It is understood that one can speak of
``propagation of light" by thinking of gauging the global $U(1)$
charge: the gravity solution allows anyway for the calculation of
the current correlators involved in the optical analysis.
Alternatively, one could maybe think about ``optical properties" of
baryon charge waves (e.g. dispersion relations).

To be concrete, in \cite{ragazzi1} it was shown that every
relativistic fluid with finite charge density $\rho$ ($n_q$ in our
solution), having a dispersive pole in the (transverse) current
retarded correlator\footnote{Clearly, here $\omega, k$ are the
frequency and wave vector.} \be\label{negref} G_T(\omega,k) \sim
\frac{i \omega {\cal B}}{i\omega - {\cal D} k^2} + P(0,k)\, , \ee
displays negative refraction, i.e. the phase and group velocities
have opposite directions, for small enough frequencies $\omega^2 <
4\pi q^2 {\cal B}$, where $q$ is the $U(1)$ gauge coupling, if the
term $P(0,k)$ can be discarded. The latter requirement was verified
by explicit gravity calculations in
\cite{ragazzi1,ragazzi2,giappo2}. In formula (\ref{negref}) one has
\be {\cal B}=\frac{\rho^2}{\varepsilon+p}\, , \qquad \qquad {\cal
D}=\frac{\eta}{\varepsilon+p}\, , \ee where $\eta$ is the shear
viscosity of the fluid. In the case at hand\footnote{We make use of
the relation $\eta/s=1/4\pi$ \cite{Kovtun:2004de}.} \be {\cal B}=
\frac{\pi {\mathrm{Vol}}(X_3) }{8{\mathrm{Vol}}(X_5)}\; N_f N_c\;
T^2 \epsilon_h \dent^2 \, , \qquad \qquad {\cal D}=\frac{1}{4\pi
T}-\frac{\epsilon_h \dent^2}{8\pi T}\left(1+\frac{1}{12}\epsilon_h
\right)  \, . \ee Thus, assuming again that the term $P(0,k)$ does
not affect the result, the charged D3-D7 Plasmas would exhibit
negative refraction for \be\label{omegacrit} \omega_c^2 < q^2
\frac{\pi^2 {\mathrm{Vol}}(X_3) }{2{\mathrm{Vol}}(X_5)}\; N_f N_c\;
T^2 \epsilon_h \dent^2 \, . \ee

\begin{figure}[htbp]
\begin{center}
\includegraphics[width=.65\textwidth]{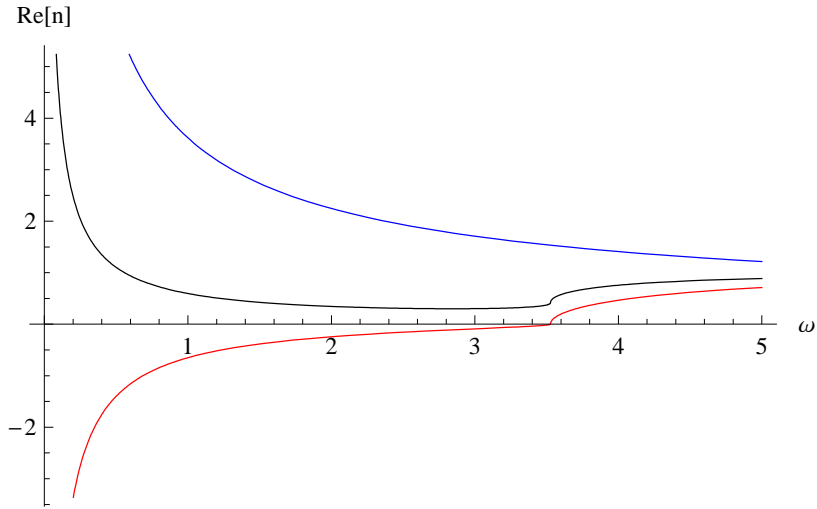}
\end{center}
\caption{The real part of the two refractive indexes $n_1$ (upper line) and $n_2$ (bottom line), with the effective one $n_{eff}$ (central line), as functions of the frequency (we are using unit temperature). The plot refers to the ${\cal N}=4$ theory with the following parameter choice: $N_c=100, N_f=20, \lambda_h=50, q=1, \dent=0.1$.}
\label{figurerefractive}
\end{figure}

Actually, in a medium with large spatial dispersion, there is the
possibility of the propagation of two ``light waves" for each
incident wave, due to the fact that the dispersion relation admits
two different solutions\footnote{Again, assuming that $P(0,k)$ is
ineffective.} \cite{ragazzi2} \be\label{ennesquared} n^2_{1,2}
=\frac{{\cal D}+\frac{i}{\omega} \pm \sqrt{\frac{16 i \pi q^2 {\cal
B}{\cal D}+\omega({\cal D}\omega-i)^2}{\omega^3}}}{2{\cal D}}\, .
\ee This effect is due to the coupling of the incident wave with
``exciton" quasi-particles in the medium, which in the dual gravity
description are quasi-normal modes. One of the two indexes should
correspond to negative refraction for small $\omega$, while the
other one is connected to the existence of an additional light wave.
The propagation of both waves in certain frequency regimes could be
inferred from the study of an ``effective refractive index" \be
n_{eff}=\frac{1+n_1 n_2}{n_1+n_2}\, . \ee

The results for the real part of the two refractive indexes and the
effective one are summarized in figure
\ref{figurerefractive}.\footnote{We thank A. Amariti, D. Forcella
and A. Mariotti for their crucial observations on the results.}
One of the two indexes displays negative refraction for $\omega <
\omega_c$, with $\omega_c$ given precisely by
(\ref{omegacrit}).\footnote{The imaginary part of one of the two
indexes, Im$[n_2]$, would be negative for $\omega < \omega_c$. This
means that we have to choose, for $\omega < \omega_c$, the negative
branch of the square root of $n_2^2$ in (\ref{ennesquared}), which
results in the negative refraction. We also find a particular value
of $\omega_p>\omega_c$ for which the two indexes exchange (i.e. the
continuous functions are made by gluing the two indexes at
$\omega_p$). This is probably an artifact of Mathematica.} The
interpolation of the effective index $n_{eff}$ from one refractive
index at small frequencies to the other one at large frequencies,
which can be seen in the plot, signals that in the intermediate
regime both light waves actually propagate.

In conclusion, the outcome has very similar features to the ones
found in the analysis of the RN black hole cases in
\cite{ragazzi1,ragazzi2,giappo2}. Thus, we seem to see the expected
features already at the level of hydrodynamics. It would be very
interesting to study the Green functions directly in the gravity
setting to understand whether this behavior is confirmed and to what
extent the hydrodynamic approximation is reliable.

\section{Conclusions}
\label{sectionconclusions}
Following the strategy initiated in  \cite{D3D7QGP}, in the present
paper we give the crank another turn,  by introducing a new
deformation  that encodes holographically the backreaction of
charged degrees of freedom at finite density. Following the same
logic we provide an  analytic solution which  is perturbative in the
deformation parameters, and explicitly write it down up to second
order in both $\epsilon_h$ and $\tilde \delta$. Moreover it is an
effective solution. This means that only IR quantities can be
reliably computed from it. In this case, the IR energy scale,
$\mu_{IR}$, is set by the temperature, and all results have to be
understood as being correct up to terms of order $\mu_{IR}/\mu_{UV}=
T/\mu_{UV}$ where $\mu_{UV}$ is an ultraviolet cutoff energy scale
above which the ultraviolet completion sets in. Still a lot of
phenomenologically interesting prediction can be made both for
thermodynamic quantities as well as for transport coefficients.  We
have shown that the thermodynamics is consistent, and we are in a
stable phase. So far, nothing can be said about phase transitions.
The solution contains two horizons, and has the potential to exhibit
an extremal solution that would allow to explore the zero
temperature - finite chemical potential axis. However such a
systematic search of the phase space in the $(T,\mu)$ plane lies
outside the domain of validity of the analytic solution presented
here. It would require the use of numerical techniques, along the
lines of the recent work \cite{DeWolfe:2010he} and  certainly
constitutes a natural continuation of the present work.

Concerning transport phenomena, we have studied the jet quenching
parameter and found that the effect of the net baryon density is
positive or negative  depending on the way one compares different
theories with and without flavor. In the present context this effect
is subdominant as compared to the one driven by the presence of
flavor. Again, the fate of this at higher finite values of the
charge density can only be established numerically, and its
importance is evident from the fact that this is one of the few
windows to phenomenology for a  rather wide class of models. It is
also worth comparing this result with the information one can obtain
from the evaluation of the drag force.

Last but not least, we have analyzed the possibility of exotic
optical phenomena in the case that the global $U(1)$ could be gauged
and treated as the electromagnetic one. Following
\cite{ragazzi1,ragazzi2}, we have checked the presence of a negative
refractive index and the propagation of additional light waves in
certain frequency regimes. Our analysis concerns the expectations
for the optical properties, based on hydrodynamic considerations. It
would be very interesting to go beyond the hydrodynamic
approximation by the explicit gravity calculation of the relevant
retarded Green functions.

\section*{Acknowledgments}
We would like to thank Paolo Benincasa,  Angel Paredes and Alfonso
V. Ramallo for useful discussions. We are indebted with Antonio
Amariti, Davide Forcella and Alberto Mariotti for their explanations
concerning the optical properties of relativistic charged fluids.
J.T. is thankful to the Front of Galician-speaking Scientists for
encouragement.

The research of F.B. and A.L.C. is supported by European Community's
Seventh Framework Programme (FP7/2007-2013) under grant agreements
n. 253937 and n. 253534, respectively. J.T. is supported by the
Netherlands Organization for Scientific Research (NWO) under the FOM
Foundation research program.

J.M. and J.T. are supported by the MICINN and  FEDER (grant
FPA2008-01838), the Spanish Consolider-Ingenio 2010 Programme CPAN
(CSD2007-00042), and the Xunta de Galicia (Conselleria de Educacion
and grant INCITE09-206-121-PR). Part of their research was done while visiting
the Erwin Schroedinger Institut Vienna and the Kavli Institute Beijing. They want to thank both  the ESI-Vienna and the  KITPC Beijing
for hospitality and finantial support (this last under grant KJCX2.YW.W10 of the Chinese
Academy of Sciences).

{ \it F. B. and A. L. C. would like to thank the Italian students,
parents, teachers and scientists for their activity in support of
public education and research.}

\appendix

\setcounter{equation}{0}
\section{Technical details}
\label{app:solut}
\subsection{Equations of motion from the ten dimensional action}

The equations of motion derived from the action $S$ as given in
(\ref{coloract}) and (\ref{flact}) are \cite{benini2008}\footnote{In
our conventions $F_p^2 = \frac{1}{p!} (F_p)^{a_1 a_2 \ldots a_p}
(F_p)_{a_1 a_2 \ldots a_p}$ (also for $H_3$). Notice moreover that
self-duality of $F_5$ has been imposed.} \beqa \nn R_{MN} -\frac12
g_{MN} R &=& \frac12 \left(
\partial_M \Phi
\partial_N \Phi - \frac12 g_{MN} \partial_P \Phi \partial^P \Phi
\right)+ \frac12 e^{2\Phi} \left( F_{1M} F_{1N} - \frac12 g_{MN}
F^2_{1}\right)
 \nn\\
 & & + \frac14e^{\Phi}\left( F_{3MPQ}F^{\ \ \, PQ}_{3N} - g_{MN} F_{3}^2\right)+ \frac14e^{-\Phi}\left( H_{3MPQ}H^{\ \ \, PQ}_{3N} - g_{MN} H_{3}^2\right)
 \nn\\
 & & + \frac{1}{96} F_{5MPQRS} F^{\, \, \ \ PQRS}_{5N}  + \frac{2\kappa^2}{\sqrt{-g}}\frac{\delta S_{fl}}{\delta g^{MN}}\, ,
\label{eomEinstein}\\
\Box \Phi &=& e^{2\Phi} F_{1}^2 + \frac12 e^{\Phi} F_{3}^2 - \frac12 e^{-\Phi} H_{3}^2 -\frac{2\kappa^2}{\sqrt{-g}} \frac{\delta S_{fl}}{\delta \Phi}\, , \label{eomDil} \\
d(e^{2\Phi} * F_1) & = & -e^{\Phi} H_3 \wedge *F_3-\frac{1}{24} \mathcal{F}^4 \wedge \Omega_2\, , \label{eomF1}\\
d(e^{\Phi} * F_3) & = & -H_3 \wedge F_5+\frac16 \mathcal{F}^3\wedge\Omega_2\, ,\label{eomF3}\\
d(*F_5) & = & d F_5 = H_3\wedge F_3-\frac12\mathcal{F}^2\wedge \Omega_2\, , \label{eomF5}\\
d(e^{-\Phi} * H_3) & = & e^{\Phi} F_1 \wedge *F_3 - F_5 \wedge F_3 + e^{\Phi} \frac{\delta}{\delta \mathcal{F}}
\sqrt{-\det (\hat{g} + e^{-\Phi/2}\mathcal{F})} \delta^{(2)}(D7)\, , \label{eomH3}
\end{eqnarray}
where the last term in the equation for $H_3$ has to be meant as an
eight-form: in particular $\delta^{(2)}(D7)$ is a short-hand
notation for the form which arises taking the derivative w.r.t.
$\cal F$ of the smeared DBI action, where $\sqrt{-\det (\hat{g} +
e^{-\Phi/2}\mathcal{F})} d^8\chi$ is replaced by \be \sqrt{-\det (g
+ e^{-\Phi/2}\mathcal{F})} |\Omega_2| d^{10}x\,.\label{sme2} \ee The
expression for the modulus of the $\Omega_2$ form will be given in a
moment.

The Bianchi identities read
 \begin{eqnarray}
\label{bianchiF1}d F_1 & = & -  g_s\Omega_2\,  , \label{biF1}\\
d F_3 & = & H_3\wedge F_1 - \mathcal{F}\wedge \Omega_2 \, ,\label{biF3} \\
d H_3 & = & 0\, .\label{biH3}
\end{eqnarray}
Finally, the Bianchi identity and EOM for the brane field
$\mathcal{F}$ are \be d \mathcal{F} = H_3 \, \label{biFF} ,\ee \be
d\left( e^{\Phi} \frac{\delta}{\delta \mathcal{F}} \sqrt{-\det
(\hat{g} + e^{-\Phi/2}\mathcal{F})} \delta^{(2)}(D7)\right) =
d(\cdots)\, , \label{eomFF} \ee where the dots represent the terms
from the WZ part of the flavor action.

After inserting the ansatz introduced  in (\ref{ansatz})   the
equations of motion (\ref{eomF1}) (\ref{eomF3}) and (\ref{eomF5}) as
well as the Bianchi identities (\ref{biF1}) and (\ref{biH3}) are
automatically satisfied. The l.h.s. of the equation of motion
(\ref{eomH3}) for $H_3$ vanishes identically upon imposing the
ansatz, and from its right hand side (explicitly rewritten taking
into account eq. (\ref{sme2}), with $|\Omega_2|=4Q_f
h^{-1/2}S^{-2}$) we can solve for $A_t'$ to obtain \be 2\pi\alpha'
A_t' = \frac{(Q_c F_{123}+8 Q_f^2 J)bFS^4}{ \sqrt{16 Q_f^2 F^2 S^4+
e^{-\Phi}(Q_c F_{123}+8 Q_f^2 J)^2 }} \, . \label{gaugefieldeom} \ee
This is an important relation, and interestingly enough, it
automatically solves the equation of motion for $A_t$. This is
because (\ref{eomFF}) is precisely the radial derivative of
(\ref{eomH3}). The Bianchi identity (\ref{biFF}) is identically
null.

The Bianchi identity for $F_3$ (\ref{biF3}) gives a nontrivial
relation between $A_t'$ and $J$. Using (\ref{gaugefieldeom}) this
can be expressed as an independent equation for $J$ which we have
added to the list below as eq. (\ref{eomJ}).

Finally we come to the Einstein-dilaton equations of motion (\ref{eomEinstein}) and (\ref{eomDil})
\beqa\label{eomansatz}
(\log b)'' &=& 4Q_f \frac{X}{Y} + 64Q_f^2  e^{-\Phi} \, b F^2  J^2 + 8Q_f^2 e^{-\Phi}  \frac{ J'^2}{F^2 S^4} + Z\, ,\\
(\log h)'' &=& - Q_c^2 \frac{b}{h^2} + 2Q_f\frac{X}{Y} + 32 Q_f^2 e^{-\Phi}\, b F^2 J^2 + 4Q_f^2 e^{-\Phi} \frac{J'^2}{F^2 S^4} + \frac{3}{2}Z \, ,\label{eomb} \\
(\log S)'' &=& - 2 b F^4 S^4 + 6 b F^2 S^6 - Q_f e^{3\Phi/2}\frac{b^2 F^3 S^{10}}{Y} - 16Q_f^2  e^{-\Phi} \, b F^2 J^2  - \frac{1}{4}Z \, , \\
(\log F)'' &=&~~ 4 b F^4 S^4 -\frac{1}{2} Q_f^2 e^{2\Phi} \, b S^8 - Q_f \frac{X}{Y} + 16Q_f^2  e^{-\Phi} \, b F^2 J^2 - 2Q_f^2 e^{-\Phi}\frac{J'^2}{F^2 S^4}\nonumber \\
&& - \frac{1}{4}Z \, , \\
(\Phi)'' &=&  Q_f^2  e^{2\Phi}\, b S^8 +2Q_f e^{3\Phi/2} \frac{b^2 F^3 S^{10}}{Y} + 2Q_f e^{\Phi/2} F S^2 Y -32 Q_f^2 e^{-\Phi} b F^2J^2 + \frac{1}{2}Z \nonumber \\
& & - 4Q_f^2 e^{-\Phi}\frac{J'^2}{F^2 S^4}\, , \\
\left[ \frac{e^{-\Phi}J'}{S^4 F^2} \right]' &=& \frac{(Q_c
F_{123}+8Q_f^2 J)bFS^4}{\sqrt{16 Q_f^2F^2 S^4+ e^{-\Phi}(Q_c
F_{123}+8Q_f^2 J)^2    }} +8e^{-\Phi}b  F^2 J \, , \label{eomJ}
\eeqa and the constraint that comes from fixing the
reparameterization gauge invariance in the radial variable reads
\beqa \nn 0 & = & - \frac12 \log' h \log' b + \frac12 (\log' h)^2 -
12(\log'S)^2 - 4\log'b \log'S\\
\nn & &  - \log'b \log'F - 8 \log'F \log'S + \frac12 \Phi'^2\\
\nn & & - \frac{b Q_c^2}{2 h^2} - 4 b F^4 S^4 + 24 b F^2 S^6 - \frac12 Z - \frac12 b e^{2\Phi} Q_f^2 S^8\\
 & & -\frac{4b^2 e^{3\Phi/2} F^3 Q_f S^{10}}{Y} - 32 b e^{-\Phi} F^2 Q_f^2 J^2 + \frac{4 e^{-\Phi} Q_f^2 J'^2}{F^2 S^4}\, . \label{apconstraint}
\eeqa
In these expressions we defined
\be
X =  (2\pi\alpha'  A'_t)^2   e^{\Phi/2}F S^2~;~~Y = \sqrt{-(2\pi\alpha'   A'_t)^2+b^2 e^\Phi F^2 S^8}~;~~Z = F_{123}^2 e^{\Phi} \, b h^2 F^2 S^8\, . \label{xyzansatz}
\ee

\subsection{Solution with explicit UV cutoff}

The solution with explicit dependence on the position of the radial
cutoff $r_*$ to first order in $\epsilon_*$ reads

\beqa
b(r)&=&1-\frac{\rhne^4}{r^4}-\frac{\tilde\delta ^2 \epsilon_*}{2} \left(2 -\frac{\rhne^4}{r^4}\right) \left(\left(1-\frac{r^2}{r_*^2}\right) \frac{\rhne^2}{r^2}-\log\left[\frac{1+\frac{\rhne^2}{r^2}}{1+\frac{\rhne^2}{r_*^2}}\right]\right) + {\mathcal{O}}(\epsilon_*^2)\, ,
 \label{solubst} \\
\frac{S(r)}{r}&=& 1+\frac{\epsilon_* }{36}\frac{ 3r_*^4-r^4-\rhne^4}{2r_*^4-\rhne^4} \nonumber \\
&& +\frac{\tilde\delta
^2 \epsilon_* }{40} \Bigg(\frac{r_*^4 \left(6-4 \frac{r^2}{\rhne^2} \right) + \frac{r_*^2}{\rhne^2} \left(12 r^4-6 \rhne^4\right) + \left(2
r^2 \rhne^2-6 r^4 \right) +\frac{\rhne^2}{r_*^2}\left(\rhne^4-2 r^4 \right)}{ 2 r_*^4-
\rhne^4}\nonumber \\
&& \qquad -\frac{4 r_*^4-\rhne^4}{ r_*^4}\frac{r_*^2}{\rhne^2}\frac{G_0(r)}{G_0(r_*)} -3 \left(1-2 \frac{r^4}{\rhne^4}\right) \log \left[\frac{1+\frac{\rhne^2}{r^2}}{1+\frac{\rhne^2}{r_*^2}}\right]\Bigg)+ {\mathcal{O}}(\epsilon_*^2)\, ,
\label{soluSst} \\
\frac{F(r)}{r} &=& 1-\frac{\epsilon_* }{36}\frac{ 3r_*^4+r^4-2\rhne^4}{2r_*^4-\rhne^4}\nonumber \\
&& +\frac{\tilde\delta ^2 \epsilon_* }{40}
\Bigg(\frac{ r_*^4\left(6 -44\frac{ r^2}{\rhne^2}+10 \frac{\rhne^2}{r^2}\right)+ r_*^2r^2\left(12\frac{r^2}{\rhne^2}-6 \frac{\rhne^2}{r^2}\right) - r^4\left(6-22 \frac{\rhne^2}{r^2}+5 \frac{\rhne^6}{r^6}\right)}{2 r_*^4 - \rhne^4}\nonumber \\
&& -\frac{\rhne^2}{r_*^2}\frac{2r^4-\rhne^4}{2 r_*^4 - \rhne^4}+\frac{16 r_*^4-4\rhne^4 }{r_*^4 }\frac{r_*^2}{\rhne^2}\frac{G_0(r)}{G_0(r_*)} -3
\left(1-2 \frac{r^4}{\rhne^4}\right)\log\left[\frac{1+\frac{\rhne^2}{r^2}}{1+\frac{\rhne^2}{r_*^2}}\right]
\Bigg) \nonumber\\
&& + {\mathcal{O}}(\epsilon_*^2)\, ,
\label{soluFst}\\
\Phi(r)&=& \Phi_*+ \epsilon_* \log \left( \frac{r}{r_*} \right) + {\mathcal{O}}(\epsilon_*^2)\, ,\\
J(r)&=&-\frac{\rhne^3 }{8 }\frac{ r_*^4-r^4}{ r_*^4+\rhne^4}+ {\mathcal{O}}(\epsilon_*)\, ,\label{soluJst}
\eeqa
with $G_0(r)=2\pi \frac{\rhne^6}{r^6}\,{}_2F_{1}\left(\frac{3}{2},\frac{3}{2},1,1-\frac{\rhne^4}{r^4}\right)$.

%%%%%%%%%%%%%%%%%%%%%%%%%%%%%%%%%%%%%%%%%%%%%%%%%%%%%%%%%%%%%%%%%%%%%%%%%%%%%

\end{document}